\documentclass[preprint,tightenlines,eqsecnum,floats,aps,amsmath,amssymb,nofootinbib,prd,showpacs,superscriptaddress]{revtex4}  
\usepackage{amsmath,amssymb,amsfonts}
\usepackage{graphicx}
\usepackage{enumerate}
\usepackage[colorlinks=true,linkcolor=blue,anchorcolor=blue,citecolor=blue,urlcolor=blue]{hyperref}
\usepackage[sort&compress]{natbib}
\usepackage{url}
\usepackage[T1]{fontenc}
\usepackage{mathptmx}
\usepackage{float}
\usepackage{booktabs}
\usepackage{CJK}
\usepackage{epstopdf}

\begin{document}
\title{Constraint on particle mass of dark matter based on XTE J1814-338 observation and  DANS model }

\author{Lin Wang}
\affiliation{School of Physics and Optoelectronics, South China University of
Technology, Guangzhou 510641, P.R. China}
\author{Dehua Wen\footnote{Corresponding author. wendehua@scut.edu.cn}}
\affiliation{School of Physics and Optoelectronics, South China University of
Technology, Guangzhou 510641, P.R. China}
\date{\today}

\begin{abstract}
The small observed radius of neutron star XTE J1814-338 (with mass of $1.21_{-0.05}^{+0.05}$ $M_{\odot}$ and radius of $7_{-0.4}^{+0.4}$ $\rm{km}$) poses a great challenge to the traditional neutron star model composed of normal nuclear matter. The dark matter admixed neutron star (DANS) model provides a clue to solve this challenge. By employing the self-interacting fermionic dark matter model, we investigate the effects of dark matter parameters including the particle mass, self-interaction strength, and mass fraction of dark matter on the properties of DANSs. Our results indicate that the joint constraints from the observation of XTE J1814-338 and the astrophysical limits on the dark matter self-interaction cross section derived from observations of different colliding galaxy clusters can provide effective constraints on the particle mass of dark matter. Specifically, to simultaneously satisfy the small radius of XTE J1814-338 and the self-interaction limit, the particle mass of self-interacting fermionic dark matter is constrained to the range from sub-GeV to a few GeV.

\end{abstract}

\pacs{97.60.Jd; 04.40.Dg; 04.30.-w; 95.30.Sf; 95.35.+d}

\maketitle

\section{Introduction}

Current understanding of the equation of state (EOS) for high-density normal nuclear matter remains subject to substantial uncertainties \cite{EOS1,EOS2,EOS3,EOS4,EOS5,EOS6}. Theoretical models have proposed a series of candidate frameworks to describe compact matter objects, including neutron-rich matter \cite{EOS5,model7}, nuclear pasta phases \cite{model3}, hyperonic matter \cite{EOS4,model4,EOS6}, and deconfined quark matter \cite{EOS4}. Neutron stars (NSs), hosting extremely dense matter in their interiors, offer a unique laboratory for probing the EOS of high-density nuclear matter. Recent observational breakthroughs related to neutron stars have substantially advanced our understanding of the nature of high-density nuclear matter. For instance, the first detection of the binary neutron star merger GW170817 in 2017 \cite{GW170817-1}, accompanied by observations of electromagnetic signals from the same source including gamma-ray burst GRB170817A and kilonova AT2017gfo \cite{GW170817-2}, represents an exciting breakthrough in gravitational wave astronomy and multi-messenger astrophysics. Analyses of the multi-messenger observations from GW170817 have constrained the tidal deformability $\Lambda_{1.4}$ of a $1.4 M_{\odot}$ neutron star to the interval $\Lambda_{1.4} = 190_{-120}^{+390}$ \cite{GW170817-3}, supporting a relatively soft EOS for high-density nuclear matter. Additionally, the Neutron Star Interior Composition Explorer (NICER) \cite{NICER1,NICER2,NICER3,NICER4,NICER5,NICER6} has accumulated simultaneous mass-radius measurements of multiple neutron stars in recent years, which also provide effective constraints on the EOS of normal nuclear matter \cite{constraints1,constraints2}.

In recent research, the pulse profile modeling of the XTE J1814-338 pulsar has inferred its mass and radius measurements: $M = 1.21_{-0.05}^{+0.05} M_{\odot} $ and $R = 7.0_{-0.4}^{+0.4}$ $\mathrm{km}$ \cite{XTE}. This observation indicates a notable discrepancy in radius relative to previous measurements of other neutron stars. In previous observations, neutron stars with mass around $1.4M_{\odot}$ tend to have radii near or above $10$ $\mathrm{km}$, for example, as inferred from the binary neutron star merger event GW170817 \cite{GW170817-3}, and pulsars PSR J0030+0451 \cite{NICER3,NICER4,PSRJ0437-4715R1,PSRJ0437-4715R2}, PSR J0437-4715 \cite{NICER5}, and PSR J0614-3329 \cite{NICER6}. Traditional neutron star models \cite{EOS1,EOS2,EOS3,EOS4,EOS5,EOS6} struggle to simultaneously satisfy both the mass and the relatively small radius observation, posing severe challenges to theoretical descriptions of neutron stars. To date, much literature have adopted different models to explain this anomalous observation. For instance, Veselsk\'{y} et al. \cite{XTED4} employed a relativistic mean-field model incorporating first-order kaon condensation ($K^{-}$ and $\bar{K}^{0}$) to soften the EOS, thereby aligning the model's mass-radius relations with the XTE J1814-338 observation. Laskos-Patkos et al. \cite{XTED5} constructed hybrid star EOS, demonstrating that descending hybrid branches can reconcile with the mass and radius observations. Zhang et al. \cite{XTEDzhang} explained the small radius observation using a new type of self-bound hybrid stars. Zhou et al. \cite{XTED7} and Dong et al. \cite{XTED7-2} utilized sound-speed transition models to simulate twin stars, fitting the small-radius branch to XTE J1814-338 while constraining parameter spaces via Bayesian analysis. Additionally, Z\"{o}llner et al. \cite{XTED6} examined nuclear core-coronal structures in neutron star models, parameterizing core structure to constrain the ranges consistent with the XTE J1814-338 observation.

Beyond adjustments to traditional EOS, interactions between dark matter and normal nuclear matter provide a novel framework for interpreting the anomalous observation of XTE J1814-338. Neutron stars, among the densest known astrophysical objects, can gravitationally capture dark matter, whose accumulation may significantly alter their mass, radius, and tidal deformability \cite{property5,property6,property1,property2,property3,property4,Sun,Lu}, and may even enable ultra-massive neutron star configurations \cite{ultramassive}. In the case of the anomalous neutron star XTE J1814-338, several studies have attempted to interpret its observational features using dark matter admixed neutron star (DANS) models. Lopes et al. \cite{XTED3,XTED3-2} utilized a single-fluid DANS model with a constant density dark matter component to deduce the microscopic and macroscopic properties of neutron stars that are consistent with the observations. Pitz et al. \cite{XTED1}, by contrast, introduced new scalar particles beyond the standard model to construct a bosonic DANS model and explored the stable parameter space consistent with observations, though at the cost of increased physical complexity. Meanwhile, Yang et al. adopted mirror dark matter \cite{XTED2} (assumed to share the same EOS as normal nuclear matter) and self-interaction bosonic dark matter \cite{XTED2-2} to soften the normal nuclear matter EOS, thereby constructing a DANS model capable of accounting for the unusual observation. The same mirror dark matter model was also employed by Jiao et al. \cite{XTED2-3} for the same purpose. Moreover, Vikiaris et al. \cite{XTEDMV} ascribe these anomalous observations to the generation of dark matter particles via neutron decay within neutron stars.

The existence of dark matter can significantly alter the structure and thermodynamic properties of neutron stars \cite{thermodynamic1,thermodynamic2}, thereby producing observable signatures in astrophysical measurements \cite{property5,property2,signatures1,signatures2,signatures3,signatures4,signatures5,gravitational1,gravitational2,gravitational3,gravitational4}. These effects establish neutron stars as astrophysical laboratories for indirectly probing the nature of dark matter \cite{property2,XTED3,XTED3-2,XTED1,XTED2,XTED2-2,Ivanytskyi,Miao,Routaray,Bell,Shakeri,Barbat,FX2,Biesdorf,Panotopoulos}. Research on constraining dark matter parameters using current and forthcoming multimessenger observations of neutron stars has attracted considerable attention \cite{parameters1,parameters2,parameters3}. In this work, we adopt a self-interacting fermionic dark matter model \cite{Fermi,Fermi2,Fermi3} to investigate DANSs. Within this model, the dark matter parameters can be constrained by the astrophysical limits on the dark matter self-interaction cross section derived from observations of different colliding galaxy clusters \cite{Markevitch,Kaplinghat,Sagunski,Loeb,wei1}. Assuming that dark matter interacts with normal nuclear matter only via gravity, we employ the two-fluid Tolman-Oppenheimer-Volkoff (TOV) equations \cite{tov1,tov2} to investigate the properties of DANSs. Our results show that the joint constraints from the observation of XTE J1814-338 and the dark matter model under Born approximation can provide effective constraints on the ranges of dark matter parameters, thereby providing a well-defined range for the dark matter particle mass.

This paper is organized as follows. In Sec. \ref{s1}, we provide a concise overview of EOS for both normal nuclear matter and dark matter, as well as the two-fluid TOV equations employed to model DANSs. In Sec. \ref{s2}, we investigate the constraints on dark matter parameters derived from the observation of XTE J1814-338, presenting the bound on the particle mass. Then we investigate the effects of the different normal nuclear matter EOSs on DANSs and present the range of the dark matter particle mass constrained under the joint constraints in Sec. \ref{s4}. Moreover, in Sec. \ref{s5}, we present a discussion of the physical implications of these results, including possible formation scenarios for high dark matter mass fractions in DANSs, a stability analysis of DANSs consistent with the small-radius observation, and a comparison with cosmological and laboratory constraints in Sec. \ref{s5}. The summary is given in Sec. \ref{s6}.

\section{The Model of dark matter admixed neutron stars} \label{s1}

\subsection{EOS of normal nuclear matter}

The APR3 EOS \cite{APR3} is employed to model the normal nuclear matter component within DANSs, thereby exploring the dark matter parameter space constrained by the joint constraints from the observation of XTE J1814-338 and the self-interaction limits. Notably, the APR3 EOS is consistent with observations from GW170817 and NICER. In addition, given the substantial uncertainties persisting in our understanding of the EOS for high-density nuclear matter, we utilize the Soft EOS \cite{Soft} and Stiff EOS \cite{Soft}, representing softer and stiffer EOS characteristics respectively, to assess the impact of different nuclear matter EOSs on DANS properties. Consequently, we adopt the Soft EOS, Stiff EOS, and APR3 EOS as representative soft, stiff, and intermediate EOS.

\subsection{Dark matter model as self-interacting fermions}
Here, we consider dark matter as self-interacting fermionic \cite{Fermi,Fermi2,Fermi3} characterized by particle mass $m_{\chi}$ and self-interacting via a repulsive Yukawa potential $V\left( r \right) =\alpha \frac{e^{-mr}}{r}$ with coupling constant $\alpha$ and mediator mass $m$. The energy density and pressure are described by
\begin{equation}
\begin{array} {c}
\varepsilon_{D}=\frac{m_{\chi}^{4}} {8 \pi^{2} ( \hbar c )^{3}} [ ( 2 x^{3}+x ) \sqrt{1+x^{2}}-\operatorname{sinh}^{-1} ( x ) ]+\frac{m_{\chi}^{4} y^{2} x^{6}} {( 3 \pi^{2} )^{2} ( \hbar c )^{3}}, \\
\end{array}
\end{equation}
\begin{equation}
\begin{array} {c}
p_{D}=\frac{m_{\chi}^{4}} {24 \pi^{2} ( \hbar c )^{3}} [ ( 2 x^{3}-3 x ) \sqrt{1+x^{2}}+3 \operatorname{sinh}^{-1} ( x ) ]+\frac{m_{\chi}^{4} y^{2} x^{6}} {( 3 \pi^{2} )^{2} ( \hbar c )^{3}}, \\
\end{array}
\end{equation}
where $x \equiv k_{F}/m_{\chi}$ with the Fermi momentum $k_{F} = \hbar c( 3 \pi^{2} n_{\chi} )^{1 / 3} $, and $y=\sqrt{2\pi \alpha m_{\chi}^2/m^2}$ is the self-interaction strength.

\begin{figure}[htbp]
    \centering
    \includegraphics[scale=0.3]{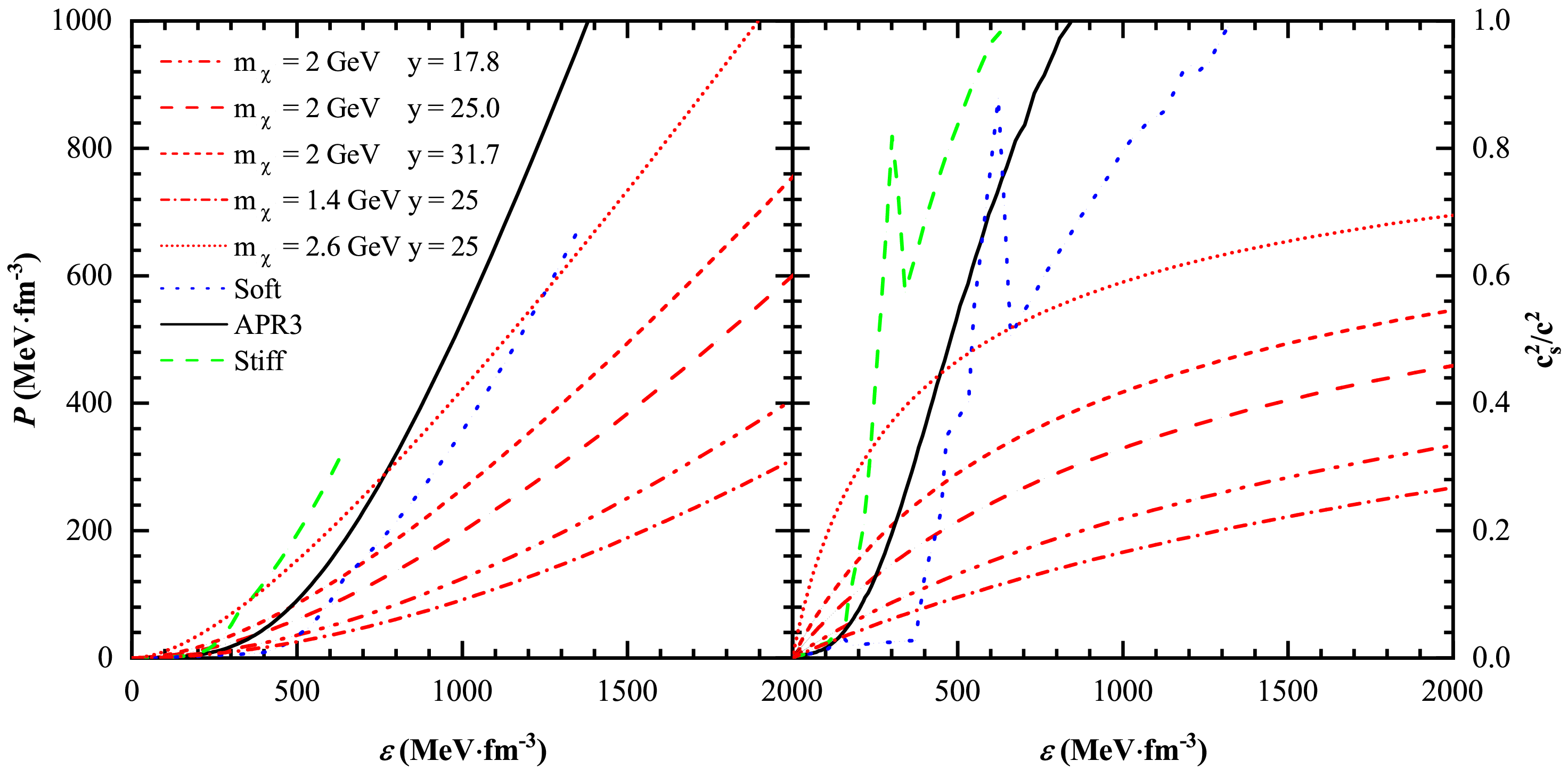}
    \caption{ The EOSs of normal nuclear model (APR3, Soft and Stiff) and dark matter model with different particle mass and self-interaction strengths, where the left and right panels show the various EOSs and the corresponding squared sound speeds, respectively. Curves are categorized as follows: red lines denote self-interacting fermions dark matter model with a series of parameters across varying several particle masses and self-interaction strengths of dark matter; blue dotted lines correspond to the Soft EOS, black solid lines depict the APR3 EOS, and green dashed lines indicate the Stiff EOS. }
    \label{EOS}
\end{figure}

Within this model, the dark matter particle mass $m_\chi$ and the self-interaction strength $y$ can be constrained by the astrophysical limits on the dark matter self-interaction cross section. Based on the observations of different colliding galaxy clusters \cite{Markevitch,Kaplinghat,Sagunski,Loeb,wei1}, this cross section is constrained to the range
\begin{equation}
\begin{array} {c}
\frac{\sigma_\chi}{m_\chi} \sim 0.1 \mathrm{-}10 \ \mathrm{cm}^2/\mathrm{g}.
\end{array}
\end{equation}
In \cite{Fermi2,dark4,Tulin,Zurek}, the Born approximation
\begin{equation}
\begin{array} {c}
\sigma_{\mathrm{Born}} = \frac{y^4}{\pi m_\chi^2}
\end{array}
\end{equation}
provides an accurate description of the self-scattering cross section for $m_\chi \lesssim 1$ $\mathrm{GeV}$ and remains valid in the limit $\alpha \to 0$ for higher masses. Within this regime, the cross section per unit mass is given by
\begin{equation}
\begin{array} {c}
\frac{\sigma}{m_\chi} = \frac{y^4}{\pi m_\chi^3}.
\end{array}
\end{equation}
Employing the conversion $1$ $\mathrm{cm}^2/\mathrm{g} = 4560/\mathrm{GeV}^3$, the observational bounds directly translate into an allowed range for the dark matter parameter
\begin{equation}
\begin{array} {c}
y^4 = \pi m_\chi^3 \frac{\sigma}{m_\chi},
\end{array}
\end{equation}
which yields
\begin{equation}
\begin{array} {c}
y \approx 6.16\, m_1^{3/4}
\end{array} \label{lower}
\end{equation}
for the lower bound, and
\begin{equation}
\begin{array} {c}
y  \approx 19.46\, m_1^{3/4}
\end{array} \label{upper}
\end{equation}
for the upper bound, where $ m_1 = m_\chi / 1\mathrm{GeV} $. Consequently, within the observationally permitted region, the dark matter EOS depends on both $m_\chi$ and the specific value of $\sigma/m_\chi$ or, equivalently, on the pair $(m_\chi, y)$ with $y$ restricted to the range derived above. The stiffness of the dark matter EOS depends critically on both the particle mass $m_{\chi}$ and self-interaction strength $y$. For a dark matter particle mass of $m_{\chi} =2$ $\rm{GeV}$, the self-interaction strength $y$ is constrained to the range $[10.34, 32.71]$ derived from the self-interaction cross-section limits via Eqs. \eqref{lower} and \eqref{upper}. As demonstrated in Fig. \ref{EOS}, increasing $y$ while holding the particle mass fixed at $m_{\chi} =2$ $\rm{GeV}$ progressively stiffens the dark matter EOS. Conversely, increasing $m_{\chi}$ at fixed $y$ significantly softens the EOS.

The astrophysical bounds on dark-matter self-interactions are derived in the low-velocity regime, whereas the dark matter inside a neutron star is highly relativistic, where the cross section is velocity-suppressed. Therefore, any overlap between the parameter region allowed by astrophysical bounds and that required by neutron-star structure is merely consistency at the level of microscopic parameters, rather than a comparison of cross sections evaluated at the same velocity.

\subsection{Two-fluid TOV equations}
In contrast to a nonrotating neutron star with a single fluid component, such as normal nuclear matter, a static two-fluid neutron star contains two individual components: normal nuclear matter and dark matter with their interaction through only gravity. These individual components require each component to be calculated independently. Therefore, we use the two-fluid TOV equations \cite{tov1,tov2}
\begin{equation}
\centering
\begin{aligned}
\frac{dp_{N}}{dr}&=-\frac{GM(r)\epsilon_{N}(r)}{c^{2}r^{2}}\left(1+\frac{4\pi r^{3}p(r)}{M(r)c^2}\right)\left(1+\frac{p_{N}(r)}{\varepsilon_{N}(r)}\right)\left(1-\frac{2GM(r)}{c^2r}\right)^{-1},\\
\end{aligned}
\end{equation}
\begin{equation}
\centering
\begin{aligned}
\frac{dp_{D}}{dr}&=-\frac{GM(r)\epsilon_{D}(r)}{c^{2}r^{2}}\left(1+\frac{4\pi r^{3}p(r)}{M(r)c^2}\right)\left(1+\frac{p_{D}(r)}{\varepsilon_{D}(r)}\right)\left(1-\frac{2GM(r)}{c^2r}\right)^{-1},\\
\end{aligned}
\end{equation}
\begin{equation}
\centering
\begin{aligned}
&\frac{dM_{N}}{dr}=\frac{4\pi r^{2}\varepsilon_{N}(r)}{c^{2}},\\
\end{aligned}
\end{equation}
\begin{equation}
\centering
\begin{aligned}
&\frac{dM_{D}}{dr}=\frac{4\pi r^{2}\varepsilon_{D}(r)}{c^{2}},\\
\end{aligned}
\end{equation}
where the subscripts $N$ and $D$ denote normal nuclear matter and dark matter, respectively. The total mass and pressure of the neutron star are expressed as the sums of two components, that is $M(r)=M_{N}(r)+M_{D}(r)$ and $p(r)=p_{N}(r)+p_{D}(r)$. The initial masses of the two fluids are zero and the central pressures of the two components are $p_{N,c}$ and $p_{D,c}$, respectively. The pressures of the two components drop until the pressures of the normal nuclear matter and dark matter vanish at different radii, thereby defining the radii of the normal nuclear matter ($R_N$) and the dark matter ($R_D$). Then the corresponding gravitational masses $M_{N}(R_N)$ and $M_{D}(R_D)$ of the two components are subsequently defined at these respective radii. The radii of the two components are compared, with the larger designated as the total radius $R$ of the neutron star. The absence of electromagnetic interaction of dark matter leads us to concentrate on investigating the radius of normal nuclear matter ($R_N$).

The two-fluid TOV equation enables a clear distinction between the dark matter radius ($R_D$) and the normal nuclear matter radius ($R_N$), making it possible to define the dark matter mass fraction $F_\chi$ \cite{parameters1,FX2,FX3} as the ratio of the gravitational mass of dark matter to the total gravitational mass $M(r)$ of the DANS, as given by the equation
\begin{equation}
F_{\chi}=\frac{M_{D}(R_{D})}{M_{D}(R_{D})+M_{N}(R_{N})}.
\end{equation}
The total gravitational mass and radius of DANS strongly depend on the value of $F_{\chi}$ \cite{parameters1,FX2,FX3}, which is an important parameter to be considered. For a given dark matter mass fraction $F_\chi$, the mass-radius relation of DANSs is uniquely determined.

Stability analyses of the two-fluid model \cite{Henriques,Leung2012,Valdez,Kain,Kain2021,Gleason,Leung2022,merger4} show that the limit of stellar stability occurs at the maximum of mass $M$ along contours of fixed particle number. Those maxima have been demonstrated \cite{Leung2012,Leung2022,merger4} to coincide with the maxima of the mass-radius ($M-R_N$) relations at fixed dark matter mass fraction $F_{\chi}$, just as in normal nuclear matter neutron stars. Configurations with higher central density (smaller radius) are unstable to induced radial oscillations, as confirmed in the cited references. In what follows we present and discuss only stable DANS configurations.

\section{Constraining dark matter parameters with the observation of XTE J1814-338 and self-interaction limit} \label{s2}
\begin{figure}[htbp]
    \centering
    \includegraphics[scale=0.35]{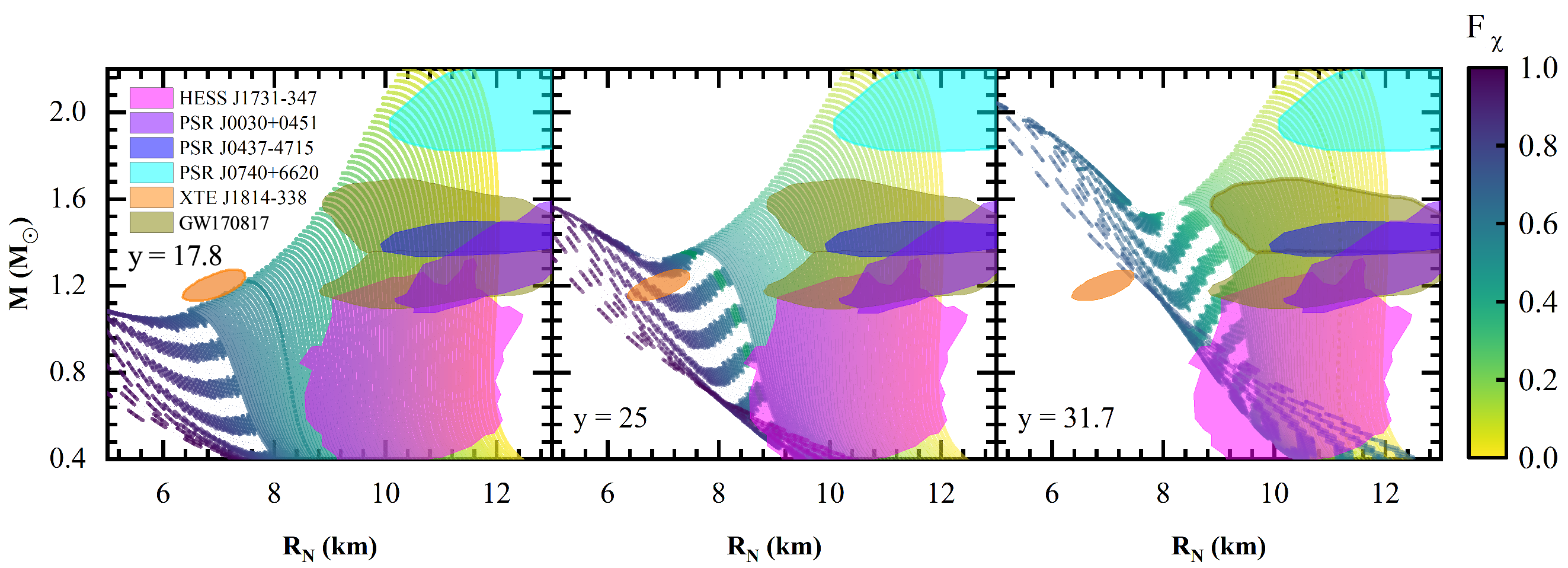}
    \caption{The relations of total mass $M$ and normal nuclear matter radius $R_N$ of DANSs under different dark matter parameters, where the normal nuclear matter EOS adopts APR3 EOS; the self-interaction strength $y$ takes values of $17.8$ (left panel), $25$ (middle panel), and $31.7$ (right panel); the particle mass of dark matter is $m_\chi=2$ $\rm{GeV}$; the dark matter mass fraction $F_{\chi}$ ranges from 0 to 1; the cyan, blue, and purple regions correspond to the mass-radius measurements of PSR J0740+6620\cite{NICER1}, PSR J0030+0451  \cite{NICER3,NICER4}, and PSR J0437-4715 \cite{NICER5}, respectively; the buff region illustrates the mass-radius observations of GW170817 \cite{GW170817-1}; the pink region depicts the mass-radius estimates inferred for the central compact object residing within the supernova remnant HESS J1731-347 \cite{HESS}; and the orange region indicates the observational mass-radius constraint derived for XTE J1814-338 \cite{XTE}.}
    \label{MR}
\end{figure}
The dark matter parameters of the DANS model are constrained by using the mass-radius observation of XTE J1814-338 and the self-interaction limit. Here, the DANS is constructed by combining the APR3 EOS for normal nuclear matter with the EOS of self-interacting fermionic dark matter characterized by the particle mass $m_\chi$ and self-interaction strength $y$. Firstly, we investigate the $M-R_N$ relations of DANS by solving the two-fluid TOV equations, which simultaneously incorporate the EOSs of both normal nuclear matter and dark matter components, with results presented in Fig. \ref{MR}. Notably, the dark matter radius $R_D$ remains electromagnetically undetectable due to its lack of electromagnetic interactions, thereby our analysis focuses on the normal nuclear matter radius $R_N$. When the radius of dark matter is smaller than the radius of normal nuclear matter, namely $R_D < R_N$, the model corresponds to a DANS with a dark matter core (solid lines). Conversely, when $R_D > R_N$, the model represents a DANS with a dark matter halo (dashed lines). For halo configurations, the dark component extending beyond nuclear matter radius $R_N$ may still affect the exterior gravitational field, which in turn influences light bending and redshift. This effect is neglected in the present work.

As illustrated in Fig. \ref{MR}, based on the $M-R_N$ relations of DANSs, it is possible to infer the dark matter mass fraction $F_\chi$  parameter space constrained by the observation of XTE J1814-338. The analysis is conducted with dark matter self-interaction strengths set to $y=17.8$, $25$, and $31.7$, and the dark matter particle mass fixed at $m_{\chi} = 2\,\rm{GeV}$. The $M-R_N$ relations form a family of curves as $F_\chi$  varies. We observe that the low mass fraction curves match most of the current observations, while satisfying the observation of XTE J1814-338 requires a higher dark matter mass fraction. For relatively weak self-interactions ($y=17.8$), the $M-R_N$ relations exhibit comparatively softness, yielding an extremely narrow range of $F_\chi$ constrained by XTE J1814-338 observation, namely $[0.66,\,0.67]$. In this case, $y=17.8$ represents the lower bound of the dark matter parameter space under the observational constraints. As the self-interaction strength $y$ increases, the corresponding $M-R_N$ curves can support larger radii and masses and still satisfy the observation of XTE J1814-338, thereby broadening the range of $F_\chi$ constrained by the observation. When the $M-R_N$ curves yield a radius exceeding the observation of XTE J1814-338, the DANS model for any value of $F_\chi$ cannot support the observation of XTE J1814-338, which can be used to define the upper bound of the dark matter parameter space as shown in the right panel of Fig. \ref{MR}. Overall, the self-interaction strength $y$ modulates the stiffness of the dark matter EOS which in turn significantly influences the $M-R_N$ relations of DANSs and dictates the range of dark matter mass fraction $F_\chi$ constrained by XTE J1814-338. As shown in Fig. \ref{MR}, the $M-R_N$ relations obtained when $y < 17.8$ fall below the observation of XTE J1814-338, whereas for $y > 31.7$ the $M-R_N$ curves exceed the observation. Based on detailed calculations with $m_{\chi} = 2$ $\rm{GeV}$ and employing the APR3 EOS for normal nuclear matter, we establish that the observation of XTE J1814-338 constrains self-interaction strength $y$ to the range $[17.8,\, 31.7]$. This interval lies entirely within the range $y \in [10.34,\,32.71]$ obtained from the self-interaction cross-section limits using Eqs. \eqref{lower} and \eqref{upper}.

\begin{figure}[htbp]
    \centering
    \includegraphics[scale=0.3]{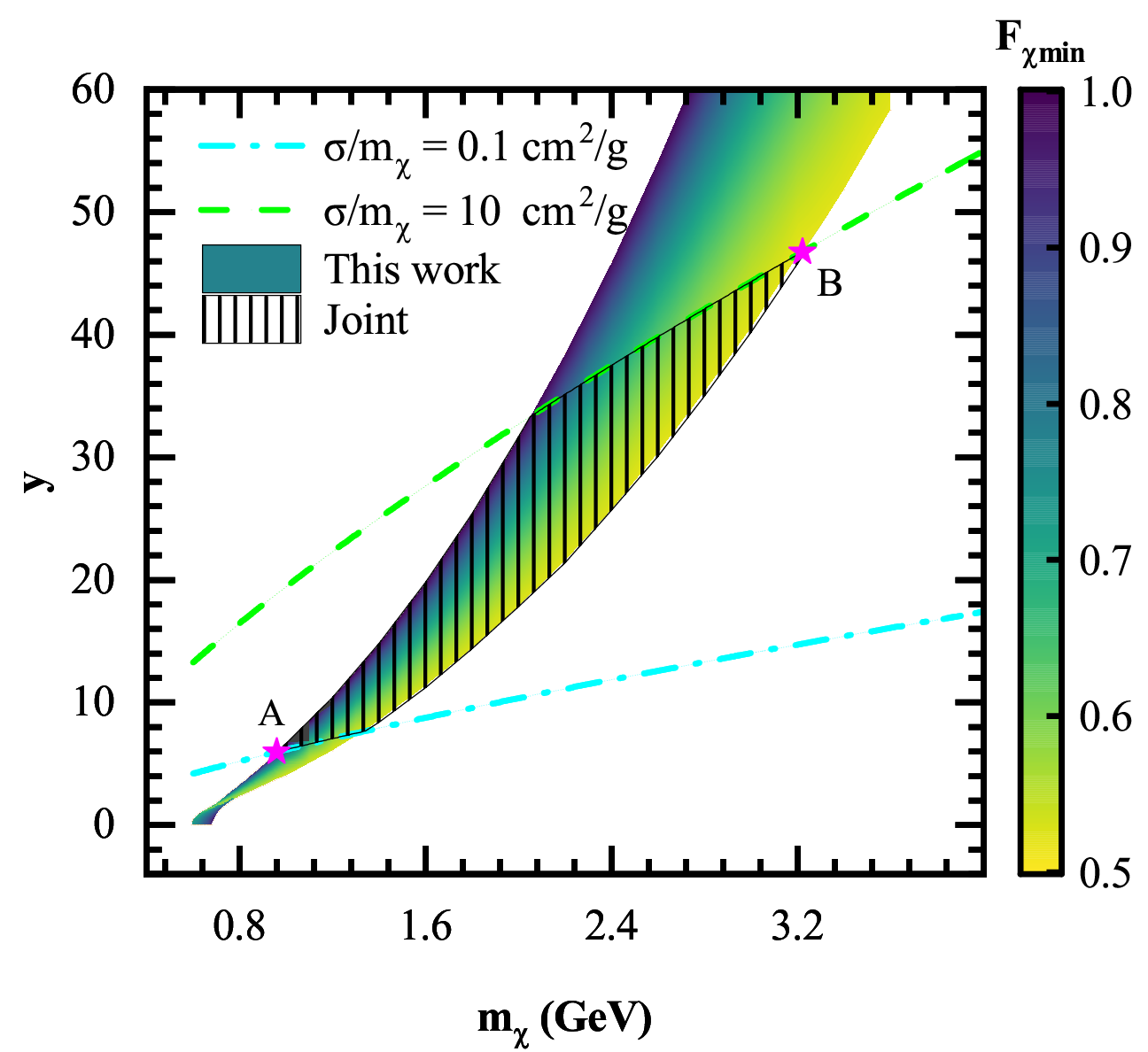}
    \caption{ Parameters space of DANSs constrained by XTE J1814-338 and dark matter model under Born approximation, where the normal nuclear matter EOS adopts APR3 EOS; cyan dot dash and green dashed lines show the lower and upper bounds from the dark matter model under Born approximation, respectively; the shaded region with the color intensity indicating the minimum dark matter mass fraction represents the parameter range allowed by the observations of XTE J1814-338, and the vertically hatched region indicates the parameter space jointly allowed by both constraints.  }
    \label{canshu}
\end{figure}

To explore the constraints imposed by the mass-radius observation of XTE J1814-338 on the parameter space of DANSs, we extend the investigation of DANSs properties to a wider range, with the particle mass $m_\chi \in [0,\,4]$ $\rm{GeV}$ and the self-interaction strength $y \in [0,\,60]$. Adopting the APR3 EOS for normal nuclear matter, we employ the joint constraints from the observation of XTE J1814-338 and astrophysical limits on the dark matter self-interaction cross section to constrain the dark matter parameter space, as illustrated in Fig. \ref{canshu}. The shaded region with the shade of color indicating the minimum dark matter mass fraction ($F_{\chi\rm{min}}$) represents the parameter range allowed by the observation of XTE J1814-338, while the vertically hatched area indicates the region further restricted by the joint constraints from the observation of XTE J1814-338 and self-interaction limits. For a given parameters set ($m_\chi, y$), the $M-R_N$ relations of DANSs constrained by the observations defines an allowed range $[F_{\chi\rm{min}}, F_{\chi\rm{max}}]$, which directly indicates that the parameters set satisfy the observational constraint. Points A and B mark the minimum and maximum dark matter particle mass under this combined constraint, respectively.  Under the mass-radius observation of XTE J1814-338, only specific combinations of particle mass and self-interaction strength yield a $M-R_N$ relation consistent with the observation. This delineates a viable parameter space for the dark matter particle mass, as shown in Fig.~\ref{canshu}. In the absence of self-interaction ($y = 0$), the observationally allowed particle mass range is $[0.593, 0.682]\,\rm{GeV}$. For stronger self-interaction $y = 10$, a higher particle mass is required, with the allowed range extending to $[1.19,\,1.52]$ $\rm{GeV}$. Thus, given the normal nuclear matter described by the APR3 EOS, the mass-radius observation of XTE J1814-338 can provide a direct constraint between $m_\chi$ and $y$, only establishing a lower limit on the dark matter particle mass $m_{\chi,\rm{min}} = 0.593\,\rm{GeV}$ which also serves as a conservative bound for the non-interacting case.

When the constraint from the dark matter self-interaction cross section is also considered, the admissible dark matter parameter space is confined to the region between the cyan and green dashed lines in Fig. \ref{canshu}. The parameter region allowed by the observation of XTE J1814-338 is then further restricted to the vertically hatched area, forming the bounded parameter space constrained by both observational bounds. Together, these constraints yield a minimum dark matter particle mass of $0.96\,\rm{GeV}$ and a maximum of $3.22\,\rm{GeV}$ under APR3 EOS for the normal nuclear matter EOS.

\section{Impact of Normal Nuclear Matter EOSs on Observational Constraints of Dark Matter Parameters} \label{s4}

\begin{figure}[htbp]
    \centering
    \includegraphics[scale=0.35]{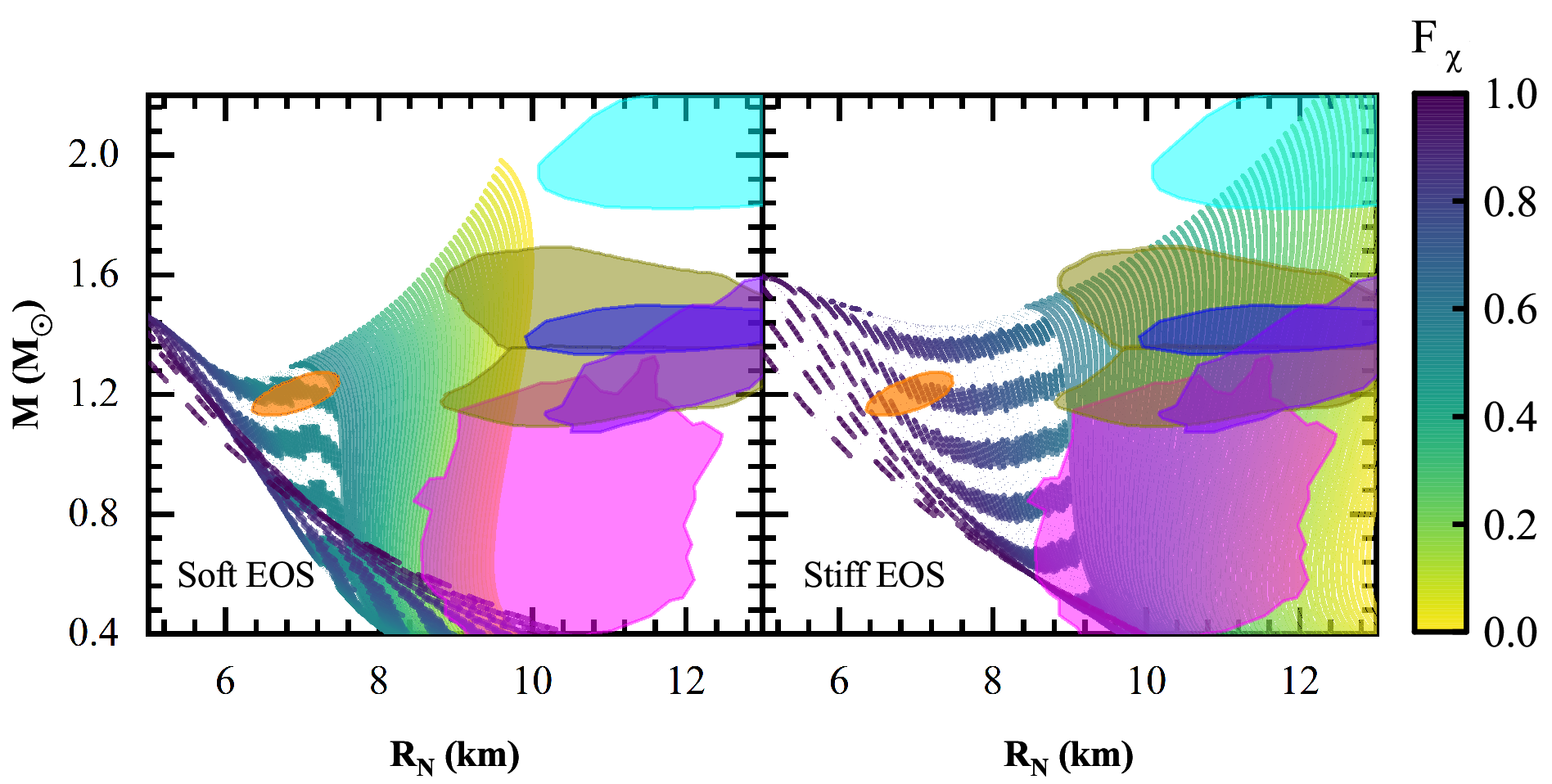}
    \caption{The relation of total mass $M$ and normal nuclear matter radius $R_N$ of DANSs, where the normal nuclear matter is described by the Soft EOS and Stiff EOS; the other parameters and shaded regions are the same with the middle panel of Fig. \ref{MR}.}
    \label{softstiff}
\end{figure}

Adopting the APR3 EOS for normal nuclear matter, we have investigated the constraints on the dark matter parameters of DANS models imposed by the XTE J1814-338 observation and the dark matter model under Born approximation. Specifically, we report the range of the dark matter particle mass constrained by these joint constraints. Given the substantial uncertainties in current understanding of the high-density nuclear matter EOS, we further incorporate diverse normal nuclear matter EOSs to examine how the choice of normal nuclear matter EOS influences DANSs under the joint constraints.

For comparison, Soft EOS and Stiff EOS were adopted as two representatives of softer and stiffer EOS, respectively. The results are presented in Fig. \ref{softstiff}. In contrast, the APR3 EOS serves as an intermediate EOS. By fixing dark matter parameters to $m_\chi=2\,\rm{GeV}$ and $y=25$, we calculate the $M-R_N$ relations of DANSs. It is also observed that only a low mass fraction is needed for the curves to remain consistent with the majority of existing observational data, while satisfying the observation of XTE J1814-338 still requires a higher dark matter mass fraction. We observe that when the Soft EOS is adopted, the observation of XTE J1814-338 intersects with the $M-R_N$ curves in their upper part, whereas under the Stiff EOS the same region overlaps with the middle portion of the curves. For the Soft EOS, the range of the dark matter mass fraction $F_\chi$ constrained by the observation of XTE J1814-338 is $[0.44,\,0.71]$, which is significantly lower than that obtained with the APR3 EOS for normal nuclear matter. This reduction arises because the Soft EOS alone cannot support neutron stars with sufficiently large masses and radii. Consequently, only a relatively small dark matter mass fraction $F_\chi$ is required to soften the $M-R_N$ curves so that they become consistent with the observation of XTE J1814-338. In contrast, the Stiff EOS, capable of supporting larger neutron star masses and radii, demands a substantially higher dark matter fraction ($F_\chi \in [0.82,\,0.95]$). These results indicate that softer normal nuclear matter EOS favors a lower $F_\chi$ range constrained by the observation, whereas stiffer EOS requires nearly critical dark matter fractions (approaching $0.99$) to compensate for the excessive stiffness of normal nuclear matter. This discrepancy indicates that the stiffness of the normal nuclear matter EOS has an impact on dark matter parameters constrained by the observation. Further calculations reveal that, for $m_\chi=2\,\rm{GeV}$, the constrained range of $y$ is $[15.8,\,37.2]$ with the Soft EOS, and $[18.4,\,32.1]$ under the Stiff EOS. Softer EOS yields a larger parameter space constrained by the observation of XTE J1814-338 for the same particle mass, which can even surpass the region $y \in [10.34,\,32.71]$ derived from astrophysical limits on the dark matter self-interaction cross section. Consequently, the self-interaction limit becomes particularly crucial for imposing further restrictions.

In order to understand the influence of different EOSs of nuclear matter on the constraint on the mass of the dark matter particle $m_\chi$, multiple EOS models are employed, and the results are summarized in Table \ref{tab1}. For example, for the Soft EOS, $m_{\chi}\in[0.837, 3.53]$ $\rm{GeV}$ is obtained, while the Stiff EOS yields $m_{\chi}\in[0.96, 3.13] $ $\rm{GeV}$. It demonstrates that the choice of the normal nuclear matter EOS has a direct but minor influence on the range of the particle mass constrained jointly by XTE J1814-338 and dark matter model under Born approximation. Specifically, to remain consistent with the observation of XTE J1814-338 and dark matter model under Born approximation, a softer normal nuclear matter EOS results in a smaller lower bound on the dark matter particle mass. In contrast, for stiffer normal nuclear matter EOSs, the required dark matter mass fraction becomes extremely large, even approaching 0.99, such that the contribution of normal nuclear matter is relatively small. Consequently, the resulting lower bound on the dark matter particle mass remains relatively unchanged. As illustrated in the left panel of Fig.~\ref{softstiff}, if an EOS is softer than the Soft EOS, it is hard to support the astronomical observations of mass-radius relations for neutron stars and is therefore excluded from the physically viable parameter space. Across the range of EOSs we examined, the lower limit on the particle mass constrained jointly by XTE J1814-338 and dark matter model under Born approximation remains in the sub-GeV range, specifically $0.837$ $\mathrm{GeV}$, while the upper limit is found to be $3.98$ $\mathrm{GeV}$.
\begin{table}[htbp]
  \centering
  \begin{tabular}{lcccccccc}
    \toprule
    \textbf{EOS type} & \textbf{Soft} & \textbf{APR3} & \textbf{Stiff} & \textbf{ALF2} & \textbf{APR} & \textbf{APR4}  & \textbf{ENG} & \textbf{DD-ME}$\delta$ \\
    \cmidrule(lr){2-9}
    $m_{\chi, \rm{min}}$ $(\rm{GeV})$ & $0.837$ & $0.960$ & $0.960$ & $0.960$ & $0.951$ & $0.960$  & $0.960$  & $0.957$  \\
    \cmidrule(lr){2-9}
    $m_{\chi, \rm{max}}$ $(\rm{GeV})$ & $3.53$ & $3.22$ & $3.13$ & $2.88$ & $3.35$ & $3.43$  & $3.15$  & $2.81$  \\
    \bottomrule
       & \textbf{MPA1} & \textbf{QMFL40}& \textbf{QMFL60}   & \textbf{QMFL80} & \textbf{SLy} & \textbf{TW99} & \textbf{WFF1} & \textbf{WFF2}\\
     \cmidrule(lr){2-9}
       & $0.958$ & $0.953$ & $0.953$ & $0.953$& $0.961$ & $0.957$ & $0.905$ & $0.948$\\
     \cmidrule(lr){2-9}
       & $3.10$ & $2.94$ & $2.89$ & $2.82$& $3.04$& $2.88$ & $3.98$ & $3.65$\\
     \bottomrule
  \end{tabular}
  \caption{Range of dark matter particle mass constrained by both the XTE J1814-338 and dark matter model under Born approximation, where the normal nuclear matter includes several EOSs, such as Soft \cite{Soft}, APR3 \cite{APR3}, Stiff \cite{Soft}, ALF2 \cite{ALF2}, APR \cite{APR3}, APR4 \cite{APR3}, ENG \cite{ENG}, DD-ME$\delta$ \cite{DDMED},  MPA1 \cite{MPA1}, QMFL40 \cite{QMFL}, QMFL60 \cite{QMFL}, QMFL80 \cite{QMFL}, SLy \cite{SLy}, TW99 \cite{TW99},  WFF1 \cite{WFF1} and WFF2 \cite{WFF1}.}
  \label{tab1}
\end{table}

\section{Discussion} \label{s5}

We employ a DANS model with self-interacting fermionic dark matter to reproduce multi-messenger astronomical observations, including gravitational waves, X-rays and thermal emissions under specific assumptions. Specifically, we constrain the relevant dark matter parameters using the joint constraints from the observation of XTE J1814-338 and dark matter model under Born approximation. Our results establish a range on the dark matter particle mass, and the analysis further shows that reproducing the XTE J1814-338 observational constraint within this framework requires a substantial dark matter mass fraction.

High dark matter mass fraction within DANSs is difficult to achieve through standard accretion processes over a typical stellar lifetime, given the relatively smooth spatial distribution of dark matter in the Galaxy. Nevertheless, several alternative formation channels could naturally produce compact stars with substantial dark matter content. In the early universe, overdense dark matter regions could act as gravitational seeds or accretion centers for baryonic matter, potentially leading to DANSs with a dominant dark matter component \cite{property6,property1,tov2,early}. In this scenario, normal nuclear matter accretes onto a pre-existing dark core rather than the neutron star accreting dark matter. Another possibility arises if a neutron star passes through regions of locally enhanced dark matter density, allowing it to capture a significant amount of dark matter during its evolution \cite{high1,property1,property5,high4,high5,high6,high7,high8}. Furthermore, stable compact objects composed primarily of asymmetric dark matter, often referred to as dark stars \cite{dark00,dark01,dark1,Fermi,Fermi2,dark4,dark5,dark6,dark7,dark8}, could serve as reservoirs or progenitors of dark matter within neutron stars. Capture from a dark star companion or merger events involving such dark compact objects may also result in neutron stars with high dark matter fractions \cite{merger1,merger2,merger3,merger4}. These scenarios indicate that high dark matter fractions are physically attainable under certain conditions but remain tightly constrained by astrophysical observations.

The formation channels discussed above indicate that a substantial dark matter component, while physically possible, is not a typical outcome of neutron star evolution. Normal neutron stars accrete at most a negligible amount of dark matter from the Galactic halo over their lifetime\cite{thermodynamic1,property5}, so other previous observations of neutron stars (PSR J0740+6620, PSR J0030+0451, PSR J0437-4715, HESS J1731-347, and GW 170817) may be explained with $F_\chi \leq 30$, and in many cases $F_\chi = 0$. Only in exceptional objects such as XTE J1814-338, the unusually small radius requires a dark-matter fraction far larger than can be supplied by ordinary halo capture, so both the required abundance and the formation probability remain uncertain, and the present work does not attempt to model their population abundance. Therefore, our model does not affect current understanding of stellar structure and evolution, nor does it conflict with current cosmological models.

Within this framework, the DANS model with self-interacting fermionic dark matter yields a mass range from sub-GeV to a few GeV, under the assumptions adopted in this work. This constraint is complementary to those from cosmology and direct detection, which probe different mass regimes. The DANS model studied in this work points to a dark matter particle mass in the GeV range, which can be tested by current and future multi-messenger observations of neutron stars.

\begin{figure}[htbp]
    \centering
    \includegraphics[scale=0.3]{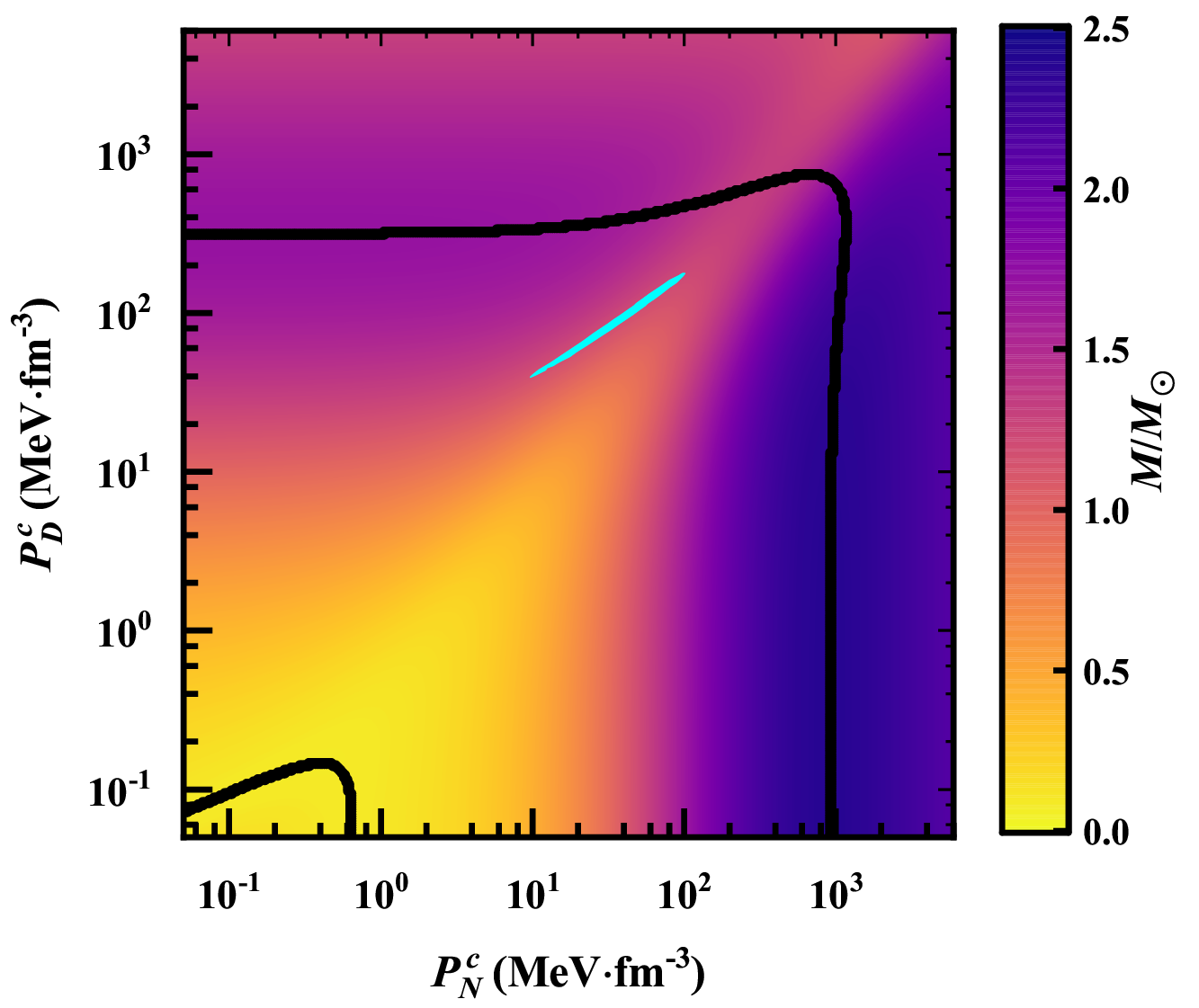}
    \caption{ Two-fluid stability analysis with self-interacting fermionic dark matter using APR3 EOS, where the shade of color  represents the total mass $M$; the black solid lines denote the stability boundary, and the cyan region indicates $M-R_N$ curves of the DANS model consistent with the observation of XTE J1814-338 as shown in  the middle panel of Fig. \ref{MR}; all other parameters are the same as those in that panel.  }
    \label{stability}
\end{figure}

Given that the small-radius configurations consistent with the observation of XTE J1814-338 imply high dark matter fractions that may lead to radial instabilities, we perform a stability analysis to assess the dynamical stability of the DANS configurations obtained from the two-fluid TOV formalism, following the criterion established in Ref. \cite{merger4}. Within this framework, the onset of unstable radial modes in a two-component system is characterized by the condition
\begin{equation}
\left( \begin{array}{c}
	\delta N_N\\
	\delta N_D\\
\end{array} \right) =\left( \begin{matrix}
	\partial N_N/\partial \epsilon _{N}^{c}&		\partial N_N/\partial \epsilon _{D}^{c}\\
	\partial N_D/\partial \epsilon _{N}^{c}&		\partial N_D/\partial \epsilon _{D}^{c}\\
\end{matrix} \right) \left( \begin{array}{c}
	\delta \epsilon _{N}^{c}\\
	\delta \epsilon _{D}^{c}\\
\end{array} \right) =0,
\end{equation}
where $N_i$ and $\epsilon _{i}^{c}$ denote the particle number and central energy density of each component, with $i = N$ for normal matter and $i = D$ for dark matter, respectively.  The particle numbers are computed via
\begin{equation}
\frac{dN_i}{dr}=4\pi r^2n_i\left( r \right) \sqrt{1-\frac{2GM\left( r \right)}{c^2r}}.
\end{equation}
Because $\epsilon _{i}^{c} \neq 0$, we have
\begin{equation}
\frac{\partial N_N}{\partial \epsilon _{cN}}\frac{\partial N_D}{\partial \epsilon _{cD}}-\frac{\partial N_N}{\partial \epsilon _{cD}}\frac{\partial N_D}{\partial \epsilon _{cN}}=0.
\end{equation}
Following the approach of previous studies \cite{XTED1,Barbat,FX2,merger4}, we present our stability results in Fig. \ref{stability}. In this figure, the black solid curves delineate the boundary between stable and unstable regimes, with stable configurations enclosed between the two stability boundaries. Notably, the $M-R_N$ curves of the DANS model discussed in this work that are consistent with the observation of XTE J1814-338, as indicated by the cyan region, fall well within the stable region, confirming their viability.

\section{Summary} \label{s6}

In this work, we construct a DANS model by incorporating self-interacting fermionic dark matter with normal nuclear matter EOS. Astrophysical limits on the dark matter self-interaction cross section are applied to obtain an initial dark matter parameter space for this dark matter model. Based on the two-fluid TOV equations, we derive the $M-R_N$ relations of DANSs across a range of dark matter parameters. We then employ the mass-radius observation of the pulsar XTE J1814-338, which is characterized by its relatively small radius, to constrain the relevant dark matter parameters. Our results indicate that the joint constraints from the observation of XTE J1814-338 and the dark matter model under Born approximation impose restrictions on the parameter space and establish a range of the dark matter particle mass. Specifically, when the APR3 EOS is adopted for normal nuclear matter, the range is inferred to be $m_{\chi} \in  [0.96, 3.22]$ $\rm{GeV}$. When we extend our analysis to several other typical EOS models of normal nuclear matter, we found that the choice of normal nuclear matter EOSs exerts a certain impact on the dark matter parameters. In conclusion, within the specific dark matter model considered here, the joint constraints from XTE J1814-338 and dark matter model under Born approximation restrict the dark matter particle mass to $[0.837, 3.98]\,\mathrm{GeV}$, i.e., from sub-GeV to a few GeV. The inferred dark-matter parameter space is driven primarily by the unusually small inferred radius of XTE J1814-338. If future analyses revise this radius, or if the source is affected by unmodeled systematics, the allowed dark-matter parameter region may become much less constraining.

Dark matter can substantially affect the macroscopic properties of neutron stars through gravitational effects or possible couplings to normal nuclear matter. These influences manifest in their mass-radius relations \cite{property5,property6}, thermal evolution \cite{thermodynamic1,thermodynamic2}, and other observable features \cite{property1,property2,property3,property4,Sun}. Validation of this framework requires coordinated multi-messenger observations. Current and next-generation gravitational-wave observatories (LIGO/Virgo/KAGRA \cite{GW170817-1,gravitational2,gravitational3,gravitational4}, and future facilities such as the Einstein Telescope \cite{ET} and Cosmic Explorer \cite{CE}) may detect signatures of dark matter through its impact on neutron stars, thereby distinguishing DANSs from normal matter neutron stars and providing dynamical constraints on dark matter properties. On the other hand, NICER and the forthcoming Athena \cite{Athena1,Athena2,Athena3} are sensitive to gravitational redshift, surface gravity, and light bending. In DANSs, these quantities are affected by the steep gravitational potentials, allowing such measurements to distinguish DANSs from normal nuclear matter neutron stars and potentially explain anomalous cooling and dark matter mediated thermal transport \cite{thermodynamic1,thermodynamic2}. Theoretical advances are needed to explore the direct interaction channels between dark matter and normal nuclear matter (such as Yukawa couplings \cite{Fermi,Fermi2,Yukawa2}) and to evaluate different dark matter candidates (bosonic dark matter \cite{bose1}, fermionic dark matter \cite{Fermi,Fermi2,Fermi3}, mirror dark matter \cite{mirror1,mirror2}, etc.), so as to achieve model self-consistency and reconcile potential observational tensions. The maturation of multi-messenger astronomy and theoretical advances should enable us to discriminate among competing models and deepen our understanding of compact-object interiors and the role of dark matter in astrophysics.

\begin{acknowledgements}
This work is supported by NSFC (Grant No. 12375144) and Guangdong Natural Science Foundation (Grant No. 2022A1515011552).
\end{acknowledgements}


\begin{thebibliography}{57}

\bibitem{EOS1}	J. M. Lattimer and M. Prakash, Astrophys. J. \textbf{550}, 426 (2001).
\bibitem{EOS2}	J. M. Lattimer, Annu. Rev. Nucl. Part. Sci. \textbf{62}, 485 (2012).
\bibitem{EOS5}	K. Hebeler, J. M. Lattimer, C. J. Pethick,  $et$ $al$., Astrophys. J. \textbf{773}, 11 (2013).
\bibitem{EOS3}	F. \"{O}zel and P. Freire, Annu. Rev. Astron. Astrophys. \textbf{54}, 401 (2016).
\bibitem{EOS4}	M. Oertel, M. Hempel, T. Kl\"{a}hn, $et$ $al$., Rev. Mod. Phys. \textbf{89}, 15007 (2017).
\bibitem{EOS6}	G. F. Burgio, H. J. Schulze,  I. Vida\~{n}a, $et$ $al$., Prog. Part. Nucl. Phys. \textbf{120}, 103879 (2021).

\bibitem{model7}J. Keller, K. Hebeler, C. J. Pethick, $et$ $al$., Phys. Rev. Lett. \textbf{132}, 232701, (2024).
\bibitem{model3}M. E. Caplan, A. S. Schneider, and C. J. Horowitz, Phys. Rev. Lett. \textbf{121}, 132701, (2018).

\bibitem{model4}L. Tolos and L. Fabbietti, Prog. Part. Nucl. Phys. \textbf{112}, 103770, (2020).


\bibitem{GW170817-1}B. P. Abbott, R. Abbott, T. D. Abbott, $et$ $al$., Phys. Rev. Lett. \textbf{119}, 161101, (2017).
\bibitem{GW170817-2}B. P. Abbott, R. Abbott, T. D. Abbott, $et$ $al$., Astrophys. J. Lett. \textbf{848}, L12, (2017).
\bibitem{GW170817-3}B. P. Abbott, R. Abbott, T. D. Abbott, $et$ $al$., Phys. Rev. Lett. \textbf{121}, 161101 (2018).

\bibitem{NICER1} T. E. Riley, A. L. Watts, A. J. Dittmann, $et$ $al$., Astrophys. J. Lett. \textbf{918}, L27 (2021).
\bibitem{NICER2} M. C. Miller, F. K. Lamb, A. J. Dittmann, $et$ $al$., Astrophys. J. Lett. \textbf{918}, L28 (2021).
\bibitem{NICER3} T. E. Riley, A. L. Watts, S. Bogdanov, $et$ $al$., Astrophys. J. Lett. \textbf{887}, L21 (2019).
\bibitem{NICER4} M. C. Miller, F. K. Lamb, A. J. Dittmann, $et$ $al$., Astrophys. J. Lett. \textbf{887}, L24 (2019).
\bibitem{PSRJ0437-4715R1} S. Vinciguerra, T. Salmi, A. L. Watts, $et$ $al$., Astrophys. J. \textbf{961}, 62 (2024).
\bibitem{PSRJ0437-4715R2} Y. Kini, L. Mauviard, T. Salmi, $et$ $al$., Astrophys. J. \textbf{1005}, 201 (2026).
\bibitem{NICER5} D. Choudhury, T. Salmi, S Vinciguerra, $et$ $al$., Astrophys. J. Lett \textbf{971}, L20 (2024).
\bibitem{NICER6} L. Mauviard, S. Guillot, T. Salmi, $et$ $al$., Astrophys. J. \textbf{995} 60 (2025).

\bibitem{constraints1} Z. Miao, A. Li, and Z.G. Dai, Mon. Not. R. Astron. Soc. \textbf{515}, 5071 (2022).
\bibitem{constraints2} J. J. Li, A. Sedrakian, and M. Alford, Phys. Rev. D \textbf{104}, L121302 (2021).
\bibitem{XTE}Y. Kini, T. Salmi, S. Vinciguerra, $et$ $al$., Mon. Not. R. Astron. Soc. \textbf{535}, 1507-1525, (2024).


\bibitem{XTED4}M. Veselsk\'{y}, V. Petousis, P.S. Koliogiannis, $et$ $al$., Phys. Rev. D \textbf{111}. L061308 (2025).
\bibitem{XTED5}P. Laskos-Patkos and Ch. C. Moustakidis, Phys. Rev. D \textbf{111}. 063058 (2025).
\bibitem{XTEDzhang}C. Zhang, J. M. Z. Pretel, and R. Xu, Phys. Rev. D \textbf{113}, 043006 (2026).

\bibitem{XTED7}T. Zhou, and C. Huang, Astrophys. J. \textbf{1002} 120 (2026).
\bibitem{XTED7-2}X. Dong, H. Shen, J. Hu, $et$ $al$., arxiv preprint arXiv:2604.21379 (2026).
\bibitem{XTED6}R. Z\"{o}llner and B. K\"{a}mpfer, Astronomy, \textbf{4}, 10 (2025).

\bibitem{property5}F. Sandin and P. Ciarcelluti, Astropart. Phys. \textbf{32}, 278 (2009).
\bibitem{property6}J. Ellis, G. H\"{u}tsi, K. Kannike, $et$ $al$., Phys. Rev. D \textbf{97}, 123007 (2018).
\bibitem{property1}M. Deliyergiyev, A. D. Popolo, L. Tolos, $et$ $al$., Phys. Rev. D \textbf{99}, 063015 (2019).
\bibitem{property2}A. E. Nelson, S. Reddy, and D. Zhou, J. Cosmol. Astropart. Phys. \textbf{07}, 012 (2019).
\bibitem{property3}Z. Rezaei, Mon. Not. R. Astron. Soc. \textbf{524}, 2015 (2023).
\bibitem{property4}H. M. Liu, J. B. Wei, Z. H. Li $et$ $al$., Phys. Dark Universe \textbf{42}, 101338 (2023).
\bibitem{Sun} H. Sun and D. Wen, Phys. Rev. D \textbf{109}, 123037 (2024).
\bibitem{Lu}W. L$\rm{\ddot{u}}$ and D. Wen. Phys. Rev. D 112, 063036 (2025).
\bibitem{ultramassive}B. K. K. Lee, M. Chu, and L. M. Lin, Astrophys. J. \textbf{922}, 242 (2021).

\bibitem{XTED3}L. L. Lopes, and  A. Issifu, Phys, Dark Universe, \textbf{48}, 101922 (2025).
\bibitem{XTED3-2} L. L. Lopes.  Astrophys. Space Sci. \textbf{370}, 79 (2025).
\bibitem{XTED1} S. L. Pitz and J. Schaffner-Bielich, Phys. Rev. D \textbf{111}, 043050 (2025).
\bibitem{XTED2} S. H. Yang, C. M. Pi, and F. Weber, Phys. Rev. D \textbf{111}, 043037 (2025).
\bibitem{XTED2-2} S.H. Yang, F. Weber., Chin. Phys. Lett. \textbf{43}, 021102 (2025).
\bibitem{XTED2-3}J. C. Jiao and C. M. Li., arxiv preprint arXiv:2606.28934 (2026).
\bibitem{XTEDMV}M. Vikiaris, V. Petousis, M Veselsky, $et$ $al$., arxiv preprint arXiv:2602. 04477 (2026).
\bibitem{thermodynamic1} J. Bramante and N. Raj, Phys. Rep. \textbf{1052}, 1 (2024).
\bibitem{thermodynamic2} F. Grippa, G. Lambiase, and T. K. Poddar, Universe \textbf{11}, 74 (2025).

\bibitem{signatures1} M. I. Gresham and K. M. Zurek, Phys. Rev. D \textbf{99}, 83008 (2019).
\bibitem{signatures2} Z. Rezaei, Astrophys. J. \textbf{835}, 33 (2017).
\bibitem{signatures3} G. Panotopoulos and I. Lopes, Phys. Rev. D \textbf{96}, 83004 (2017).
\bibitem{signatures4} A. De Lavallaz and M. Fairbairn, Phys. Rev. D \textbf{81}, 123521 (2010).
\bibitem{signatures5} O. Ivanytskyi, Phys. Rev. D \textbf{102}, 63028 (2020).

\bibitem{gravitational1} J. Ellis, A. Hektor, G H\"{u}tsi, $et$ $al$., Phys. Lett. B \textbf{781}, 607 (2018).
\bibitem{gravitational2} H. R. R\"{u}ter, V. Sagun, W. Tichy, $et$ $al$., Phys. Rev. D \textbf{108}, 124080 (2023).
\bibitem{gravitational3} M. Bezares and D. Vigan\`{o}, and C. Palenzuela, Phys. Rev. D \textbf{100}, 044049 (2019).
\bibitem{gravitational4} C. J. Horowitz and S. Reddy, Phys. Rev. Lett. \textbf{122}, 071102 (2019).



\bibitem{Ivanytskyi} O. Ivanytskyi, V. Sagun, and I. Lopes, Phys. Rev. D \textbf{102}, 063028 (2020).
\bibitem{Miao} Z. Q. Miao, Y. F. Zhu, A. Li,  $et$ $al$., Astrophys. J. \textbf{936}, 69 (2022).
\bibitem{Routaray} P. Routaray, S. R. Mohanty, H. C. Das, $et$ $al$., J. Cosmol. Astropart. Phys. \textbf{10}, 073 (2023).
\bibitem{Bell} N. F. Bell, G. Busoni, S. Robles, $et$ $al$., J. Cosmol. Astropart. Phys. \textbf{04}, 006 (2024).
\bibitem{Shakeri} S. Shakeri and D. R. Karkevandi, Phys. Rev. D \textbf{109}, 043029 (2024).
\bibitem{Barbat} M. F. Barbat, J. Schaffner-Bielich, and L. Tolos, Phys. Rev. D \textbf{110}, 023013 (2024).
\bibitem{FX2}S., Domenico, and I. Bombaci. J. High Energy Astrophys. \textbf{45}, 371 (2025).
\bibitem{Biesdorf} C. Biesdorf, J. Schaffner-Bielich, and L. Tolos, Phys. Rev. D \textbf{111}, 083038 (2025).
\bibitem{Panotopoulos} G. Panotopoulos, A. Rinc$\rm{\acute{o}}$n, and I. Lopes, Phys. Dark Universe \textbf{49}, 101972 (2025).


\bibitem{parameters1}D. R. Karkevandi, S. Shakeri, V. Sagun, $et$ $al$., Phys. Rev. D \textbf{105}, 023001 (2022).
\bibitem{parameters2}S. Bhattacharya, B. Dasgupta, R. Laha, $et$ $al$., Phys. Rev. Lett. \textbf{131}, 091401 (2023).
\bibitem{parameters3}S. Banerjee, S. Bera, and D. F. Mota, J. Cosmol. Astropart. Phys. \textbf{03}, 041 (2023).



\bibitem{Fermi} G. Narain, J. Schaffner-Bielich, and I. N. Mishustin, Phys. Rev. D \textbf{74}, 063003 (2006).
\bibitem{Fermi2} C. Kouvaris and N. G. Nielsen, Phys. Rev. D \textbf{92}, 063526 (2015).
\bibitem{Fermi3} S. Mukhopadhyay, D. Atta, K. Imam, $et$ $al$., Eur. Phys. J. C \textbf{77}, 440 (2017); \textbf{77}, 553(E)(2017).

\bibitem{Markevitch} M. Markevitch, A. H. Gonzalez, D. Clowe, $et$ $al$., Astrophys. J. \textbf{606}, 819 (2004).
\bibitem{Kaplinghat} M. Kaplinghat, S. Tulin,  and H. B. Yu, Phys. Rev. Lett. \textbf{116}, 041302 (2016).
\bibitem{Sagunski} L. Sagunski, S. Gad-Nasr, B. Colquhoun, $et$ $al$., Astropart. Phys. \textbf{01}, 024 (2021).
\bibitem{Loeb} A. Loeb, Astrophys. J. Lett. \textbf{929}, L24 (2022).
\bibitem{wei1} H. M. Liu, J. B. Wei, Z. H. Li, $et$ $al$., Phys. Rev. D \textbf{110}, 023024 (2024).

\bibitem{tov1}E. Giangrandi, V. Sagun, O. Ivanytskyi, $et$ $al$., Astrophys. J. \textbf{953}, 115 (2023).
\bibitem{tov2}P. Ciarcelluti, and F. Sandin., Phys. Lett. B \textbf{695}, 19 (2011).

\bibitem{APR3}A. Akmal and V. R. Pandharipande, Phys. Rev. C \textbf{156}, 2261 (1997).
\bibitem{Soft}K. Hebeler, J. M. Lattimer, C. J. Pethick, $et$ $al$., Astrophys. J. \textbf{773}, 11 (2013).



\bibitem{Tulin} S. Tulin, H. B. Yu, and K. M. Zurek, Phys. Rev. D \textbf{87}, 115007 (2013).
\bibitem{Zurek} K. M. Zurek, Phys. Rep. \textbf{537}, 91 (2014).
\bibitem{dark4} A. Maselli, P. Pnigouras, N. G. Nielsen, $et$ $al$., Phys. Rev. D \textbf{96}, 023005 (2017).

\bibitem{FX3}M. Collier, D. Croon1, and R. K. Leane, Phys. Rev. D \textbf{106}, 123027 (2022).


\bibitem{Henriques} A. B. Henriques, A. R. Liddle, and  R. G. Moorhouse, Phys. Lett. B \textbf{251}, 511 (1990).
\bibitem{Leung2012} S. C. Leung, M. C. Chu, and L. M. Lin, Phys. Rev. D \textbf{85}, 103528 (2012).
\bibitem{Valdez} S. Valdez-Alvarado, C. Palenzuela,  D. Alic, $et$ $al$., Phys. Rev. D \textbf{87}, 084040 (2013).
\bibitem{Kain} B. Kain, Phys. Rev. D \textbf{102}, 023001 (2020).
\bibitem{Kain2021} B. Kain, Phys. Rev. D \textbf{103}, 043009 (2021).
\bibitem{Gleason} T. Gleason, B. Brown, and B. Kain, Phys. Rev. D \textbf{105}, 023010 (2022).
\bibitem{Leung2022} K. L. Leung, M. C. Chu, and L. M. Lin, Phys. Rev. D \textbf{105}, 123010 (2022).
\bibitem{merger4} M. Hippert, E. Dillingham, H. Tan, $et$ $al$., Phys. Rev. D \textbf{107}, 115028 (2023).

\bibitem{HESS}V. Doroshenko, V. Suleimanov, G. P\"{u}hlhofer, $et$ $al$., Nat. Astron. \textbf{6}, 1444-1451 (2022).


\bibitem{ALF2} M. Alford, M. Braby, Astrophys. J. \textbf{629}, 969 (2005).
\bibitem{ENG} L. Engvik, E. Osnes,  M. Hjorth-Jensen, $et$ $al$., Astrophys. J. \textbf{469}, 794 (1996).
\bibitem{DDMED} X. Roca-Maza, X. Vi$\rm{\tilde{n}}$as, M. Centelles $et$ $al$., Phys.  Rev.  C \textbf{84}, 054309 (2011).
\bibitem{MPA1} H. Muther, M. Prakash, and T. L. Ainsworth, Phys. Lett. B \textbf{199}, 469 (1987).
\bibitem{QMFL} Z. Y. Zhu, A. Li, , J. N. Hu, $et$ $al$., Phys.  Rev.  C \textbf{99}, 025804 (2019).
\bibitem{SLy} F. Douchin and P. Haensel, Astron. Astrophys. \textbf{380}, 151 (2001).
\bibitem{TW99} S. Typel and H.H. Wolter, Nucl. Phys. A \textbf{656}, 331 (1999).
\bibitem{WFF1} R. B. Wiringa, V. Fiks, and A. Fabrocini, Phys. Rev. C \textbf{38}, 1010 (1988).



\bibitem{early} I. Goldman, R. N. Mohapatra, S. Nussinov, $et$ $al$., Phys. Lett. B \textbf{725}, 200 (2013).

\bibitem{high1} A. Li, F. Huang, and R. X. Xu, Astropart. Phys. \textbf{37}, 70 (2012).
\bibitem{high4} A. Del Popolo, M. Le Delliou, and M. Deliyergiyev, Universe \textbf{6}, 222 (2020).
\bibitem{high5} A. Del Popolo, M. Deliyergiyev, and M. Le Delliou, Phys. Dark Universe \textbf{30}, 100622 (2020).
\bibitem{high6} S. Profumo, K. Sigurdson, and M. Kamionkowski, Phys. Rev. Lett. \textbf{97}, 031301 (2006).
\bibitem{high7} V. S. Berezinsky, V. I. Dokuchaev, and Y. N. Eroshenko, J. Cosmol. Astropart. Phys. \textbf{11}, 059 (2013).
\bibitem{high8} V. S. Berezinsky, V. I. Dokuchaev, and Y. N. Eroshenko, Phys. Usp. \textbf{57}, 1 (2014).

\bibitem{dark00}D. Spolyar1, K. Freese, and P. Gondolo, Phys. Rev. Lett. \textbf{100}, 051101 (2008).
\bibitem{dark01}C. Ilie,  J. Paulin,  and K. Freese, Proc. Natl. Acad. Sci. \textbf{120}, e2305762120 (2023).
\bibitem{dark1}J. Eby, C. Kouvaris, N. G. Nielsen, $et$ $al$., J. High Energ. Phys. \textbf{2016}, 28 (2016).

\bibitem{dark5}L. Visinelli, S. Baum, J. Redondo, $et$ $al$., Phys. Lett. B \textbf{777}, 64-72 (2018).
\bibitem{dark6}P. H. Chavanis., Phys. Rev. D \textbf{98}, 023009 (2018).
\bibitem{dark7}M. Cassing, A. Brisebois, M. Azeem,$et$ $al$., Astrophys. J. \textbf{944} 130 (2023).
\bibitem{dark8}D. Croon1, S. Ipek, and D. McKeen, Phys. Rev. D \textbf{107}, 063012 (2023).

\bibitem{merger1} K. Clough, T. Dietrich, and J. C. Niemeyer, Phys. Rev. D \textbf{98}, 083020 (2018).
\bibitem{merger2} T. Dietrich, F. Day, K. Clough, $et$ $al$., Mon. Not. R. Astron. Soc. \textbf{483}, 908 (2019).
\bibitem{merger3} R. F. Diedrichs, N. Becker, C Jockel, $et$ $al$., Phys. Rev. D \textbf{108}, 064009 (2023).





\bibitem{ET} M. Maggiore, C. V. D. Broeck, N. Bartolo, $et$ $al$., J. Cosmol. Astropart. P.  \textbf{2020}, 50 (2020).
\bibitem{CE} E. D. Hall. Galaxies \textbf{10}(4), 90 (2022).


\bibitem{Athena1} A. Majczyna, J. Madej, M. Nale\.{z}yty, et al Astrophys. J. \textbf{888}, 123 (2020).
\bibitem{Athena2} R. Cassano, R. Fender, C. Ferrari, $et$ $al$., arXiv:1807. 09080 (2018).
\bibitem{Athena3} X. Barcons, D. Barret, A. Decourchelle, $et$ $al$., arXiv:1207. 2745 (2012).





\bibitem{Yukawa2} M. I. Gresham1 and K. M. Zurek. Phys. Rev. D \textbf{99}, 083008 (2019).
\bibitem{bose1}M. Colpi, S. L. Shapiro, and I. Wasserman, Phys. Rev. Lett. \textbf{57}, 2485 (1986).

\bibitem{mirror1} R. Foot, H. Lew, R. R. Volkas, Phys. Lett. B \textbf{272}, 67 (1991).
\bibitem{mirror2} Z. Berezhiani, R. Biondi, M. Mannarelli, $et$ $al$., Eur. Phys. J. C \textbf{81}, 1036 (2021).




\end{thebibliography}
\end{document}